\documentclass[10pt,aps,prl,twocolumn,preprintnumbers]{revtex4-1}

\usepackage{verbatim}
\usepackage{amsmath}
\usepackage{amsfonts}
\usepackage{amssymb}
\usepackage{bbm}
\usepackage{latexsym}
\usepackage{hhline}
\usepackage{multirow}

\usepackage{color}
\usepackage{graphicx}
\usepackage[table]{xcolor}
\usepackage{cancel}
\usepackage{tabu} 

\usepackage{capt-of}

\usepackage{hyperref}

\usepackage[colorinlistoftodos]{todonotes} 

\definecolor{LightPurple}{rgb}{1,0.85,1}
\definecolor{LightBlue}{rgb}{0.85,1,1}
\definecolor{LightRed}{rgb}{1,0.75,0.75}
\definecolor{LightGreen}{rgb}{0.85,1,0.85}

\renewcommand{\d}{\mathrm{d}}

\DeclareMathSymbol{\mg}{\mathrel}{symbols}{"1D}

\newcommand{\ga}{\alpha}

\newcommand{\gd}{\delta}
\renewcommand{\ge}{\epsilon}

\newcommand{\gm}{\mu}

\newcommand{\gl}{\lambda}
\newcommand{\gr}{\rho}
\newcommand{\gth}{\theta}

\newcommand{\gs}{\sigma}

\newcommand{\gp}{\pi}
\newcommand{\gps}{\psi}

\newcommand{\gch}{\chi}
\newcommand{\gF}{\Phi}

\newcommand{\gL}{\Lambda}
\newcommand{\gS}{\Sigma}

\newcommand{\gPs}{\Psi}

\newcommand{\cZ}{{\cal Z}}
\newcommand{\ui}{{\underline i}}

\newcommand{\tW}{{\widetilde W}}

\newcommand{\dga}{{\dot \alpha}}
\newcommand{\dgb}{{\dot\beta}}

\newcommand{\tr}{\text{tr}}

\newcommand{\Id}{\text{\small 1}\hspace{-3.5pt}\text{1}}

\newcommand{\ra}{\rightarrow}

\newcommand{\der}{\partial}

\newcommand{\scp}{\scriptstyle}

\newcommand{\beq}{\begin{equation}}
\newcommand{\eeq}{\end{equation}}
\newcommand{\barr}{\begin{array}}
\newcommand{\earr}{\end{array}}
\newcommand{\equ}[1]{\begin{gather} #1 \end{gather}}

\newcommand{\enums}[1]{\begin{enumerate} #1 \end{enumerate}}

\newcommand{\arry}[2]{\begin{array}{#1} #2 \end{array}}

\newcommand{\non}{\nonumber}
\newcommand{\sfrac}[2]{\mbox{$\frac{#1}{#2}$}}
\newcounter{oldcounter}

\newcommand{\bx}{{\bar x}}

\newcommand{\bD}{{\overline D}}

\newcommand{\bgr}{{\bar\rho}}
\newcommand{\bgth}{{\bar\theta}}

\newcommand{\bgF}{{\overline\Phi}}

\newcommand{\bgL}{{\overline\Lambda}}

\newcommand{\tgd}{{\tilde \delta}}

\newcommand{\Intr}{\mathbb{Z}}

\newcommand{\ba}[2]{\[\begin{array}{#2}\label{#1}}
\newcommand{\ea}{\end{array}\]}
\newcommand{\be}{\begin{equation}}
\newcommand{\ee}{\end{equation}}
\newcommand{\bea}{\begin{eqnarray}}
\newcommand{\eea}{\end{eqnarray}}

\newcommand{\brkt}[2]{\left[{\scp #1}\atop{\scp #2}\right]}

\newcommand{\rep}[1]{\mathbf{#1}}

\begin{document}

%
%
\title{Twisted superspace: 
Non-renormalization and fermionic symmetries in certain (heterotic string inspired) non-supersymmetric field theories}

\date{\today}
\author{ Stefan \surname{Groot Nibbelink}}
\email[]{Groot.Nibbelink@physik.uni-muenchen.de}
\author{ Erik \surname{Parr}}
\email[]{Erik.Parr@physik.uni-muenchen.de}
\affiliation{Arnold Sommerfeld Center for Theoretical Physics, Ludwig-Maximilians-Universit\"at M\"unchen, Theresienstra\ss e 37, 80333 M\"unchen, Germany}
\begin{abstract}

Inspired by the tachyon-free non-supersymmetric heterotic SO(16)$\times$SO(16) string we consider a special class of non-supersymmetric field theories: Those that can be obtained from supersymmetric field theories by supersymmetry breaking twists.
We argue that such theories, like their supersymmetric counter parts, may still possess some fermionic symmetries as left-overs of the super gauge transformations and have special one-loop non-renormalization properties due to holomorphicity.
In addition, we extend the supergraph techniques to these theories to calculate some explicit supersymmetry-breaking corrections.
\end{abstract}
%
\preprint{LMU-ASC 21/16} 
%
%
\maketitle

\section{Introduction}

Supersymmetry (SUSY) is a very powerful proposed symmetry between bosons and fermions. 
Any SUSY extension of a field theory has an effectively doubled spectrum. E.g., the minimal SUSY extension of the standard model (SM) of particle physics, the so-called MSSM, contains gauginos, squarks and sleptons as the superpartners of the SM gauge fields, quarks and leptons. Since, so far no experimental evidence for such superpartners has been found, one could take the point of view, that SUSY is simply irrelevant for physics beyond the SM.  In this Letter we point out that there exists a class of non-SUSY field theories that share quite a few of the amazing properties of SUSY models.

Surprisingly, the existence of such non-SUSY theories can be motivated by string theory: There exist perfectly well-defined (e.g. modular invariant, anomaly-free and tachyon-free) non-SUSY string theories, like the heterotic SO(16)$\times$SO(16) string~\cite{AlvarezGaume:1986jb,Dixon:1986iz}, which can be taken as the starting point for non-SUSY heterotic model building~\cite{Faraggi:2007tj,Abel:2015oxa,Ashfaque:2015vta,Blaszczyk:2015zta,Nibbelink:2015vha}.
The non-SUSY SO(16)$\times$SO(16) theory is closely related to the SUSY E$_8\times$E$_8$ and SO(32) strings, as the following two key observations show~\cite{Dixon:1986iz}: 
\enums{
\item[\text{ A.)}] The string partition functions of the SUSY E$_8\times$E$_8$ and the non-SUSY SO(16)$\times$SO(16) string theories are identical up to some generalized torsion phases. 
\item[\text{B.)}] The SO(16)$\times$SO(16) string can be obtained from either the SUSY SO(32) or E$_8\times$E$_8$ by appropriately chosen SUSY-breaking (\cancel{SUSY}) twists. 
}
The former we take as a strong indication that also the non-SUSY heterotic string should possess a number of striking features. 
(Further support for this is provided by the fact that the vacuum energy is finite~\cite{AlvarezGaume:1986jb,Kutasov:1990sv},  
misaligned supersymmetry features~\cite{Dienes:1994np,Dienes:1995pm} 
and computation of gauge thresholds~\cite{Angelantonj:2014dia,Angelantonj:2015nfa}.) 
To exhibit and generalize them we take inspiration from point B.) and construct non-SUSY models from SUSY field theories by applying \cancel{SUSY} twists to them.

\subsection{Letter overview}  

Since our main source of inspiration is provided by the non-SUSY tachyon-free SO(16)$\times$SO(16) string, we recall some of its properties, like its relation to the heterotic SUSY strings. Next, we introduce the 4D N=1 superspace formalism to provide us with a convenient framework to describe possible \cancel{SUSY} twists. In particular, we argue that the surviving fermionic parts of the super gauge transformations can be interpreted as novel fermionic symmetries. 
Moreover, we show that supergraph methods can be easily extended to non-SUSY theories by introducing \cancel{SUSY}-twist compatible superspace delta functions.  
In addition, we show that the superpotential is still a convenient concept as it does not renormalize at the one-loop level. However, already at this order the fact that such theories are not SUSY is communicated: ``soft''-SUSY breaking operators are induced.

\section{The heterotic SO(16)$\times$SO(16) string}

The massless 10D spectrum of the non-SUSY tachyon-free heterotic SO(16)$\times$SO(16) is given by~\cite{Dixon:1986iz}:
\vspace{-3ex} 
\begin{center} 
  \setlength{\tabcolsep}{0pt}
    \renewcommand{\arraystretch}{1.4}
\begin{tabular}{|c|l|}
\hline
\textbf{Fields} & \textbf{\quad Space-time interpretation}
\\ \hline\hline 
 \cellcolor{LightBlue} \,$G_{MN}, B_{MN}, \phi$~ 
 &  \cellcolor{LightBlue} Graviton, two-form, dilaton~
\\[-1pt]
 \cellcolor{LightBlue}\,$A_M^{A'},~A_M^{A''}$~ 
 &  \cellcolor{LightBlue} Gauge fields in the $(\mathbf{120},\mathbf{1})+(\mathbf{1},
\mathbf{120})$~  
\\ \hline\hline  
 \cellcolor{LightPurple}$\Psi_L^{\,a'},~ \Psi_L^{\,a''}$ &  \cellcolor{LightPurple}~Spinors in the $(\mathbf{128},\mathbf{1})+(\mathbf{1},
\mathbf{128})$
\\[-1pt]
 \cellcolor{LightPurple}
$\gPs_R^{\,m'n''}$ &  \cellcolor{LightPurple}~Cospinors in the $(\mathbf{16},\mathbf{16})$
\\ \hline
\end{tabular}
  \renewcommand{\arraystretch}{1}
\end{center}
Here, the single and double primes distinguish vector ($m',m''=1,\ldots, 16$), spinor  ($a', a''=1,\ldots,128$) and anti-symmetric adjoint 
($A' =[m',n'],
 A''=[m'',n'']
 $) indices of the first and second $SO(16)$ factor, respectively, and $M,N$ are 10D Lorentz indices.
Clearly, this spectrum is not supersymmetric: For example, there are no fermions in the gravitational sector and the gauginos of the SO(16)$\times$SO(16) gauge fields are absent.

\subsection{A.\ Partition functions of the SO(16)$\times$SO(16) and E$_8\times$E$_8$ theories}

The string partition functions encode the full string spectra. Following~\cite{Blaszczyk:2014qoa}, let ${\mathbf Z}_8^x$ denote the real partition function corresponding to the 10D  
coordinate fields in light-cone gauge and $\widehat {\mathbf Z}_d$ the modular covariant holomorphic partition functions associated to worldsheet fermions with Ramond and Neveu-Schwarz boundary conditions associated to the three spin structures $s,t,u =0,1$. 

Then, the non-SUSY heterotic SO(16)$\times$SO(16) string has a one-loop torus partition function given by 
\begin{equation} \label{SO16_partition}
{\mathbf Z}_\text{SO(16)$^2$}
= 
- 
\frac18 \sum
T \cdot 
{\mathbf Z}_8^x \cdot 
\widehat {\mathbf Z}_4\brkt{s}{s'}\cdot 
\overline{ \widehat {\mathbf Z}_8\brkt{t}{t'}} \cdot 
\overline{ \widehat {\mathbf Z}_8\brkt{u}{u'}}~,
\end{equation}
which is identical to that of the SUSY heterotic E$_8\times$E$_8$ string
\begin{equation} \label{E8_partition} 
{\mathbf Z}_{\text{E}_8^2} = 
- 
\frac18 \sum
\phantom{T \cdot}  
{\mathbf Z}_8^x \cdot 
\widehat {\mathbf Z}_4\brkt{s}{s'} \cdot 
\overline{ \widehat {\mathbf Z}_8\brkt{t}{t'}} \cdot 
\overline{ \widehat {\mathbf Z}_8\brkt{u}{u'}}~,
\end{equation}
except that \eqref{SO16_partition} has additional generalized torsion phases, $T=T_\text{torsion} \cdot T_\text{chiral}$ with 
\begin{equation} \label{TorsionPhases}
\arry{lcl}{ 
T_\text{torsion} &=& \phantom{ - }
(-)^{s t' - s' t} (-)^{s u' - s' u}  (-)^{t u' - t' u}~, 
\\[1ex] 
T_\text{chiral} &=& 
- (-)^{s's+s'+s} (-)^{t't + t' +t} (-)^{u'u + u' + u }~. 
}
\end{equation} 
They modify the GSO projections such that all SUSYs are broken~\cite{Dixon:1986iz,Blaszczyk:2014qoa}.

\subsection{Visualizing lattice sums}

Torus partition functions can be represented as sums over the $d$-dimensional root ({\bf R}$_d$), vector ({\bf V}$_d$), spinor ({\bf S}$_d$) and cospinor ({\bf C}$_d$) lattices~\cite{Blaszczyk:2014qoa}. The lattice sums of the heterotic E$_8\times$E$_8$ theory can be visualized as: 
\begin{center}
\includegraphics[scale=1]{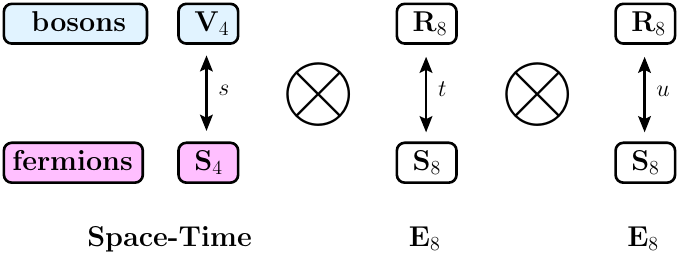}
\end{center}
The two tensor products imply that there are eight sectors in total. 
Flipping the spin structures $t$ or $u$ takes us from the SO(16) root lattice {\bf R}$_8$ to the spinorial lattice {\bf S}$_8$, which combined form the E$_8$ root lattice. The spin structure $s$ generates the E$_8\times$E$_8$ SUSY spectrum: flipping the spin structure $s$ interchanges bosons and fermions.   

Similarly, one can visualize the partition function of the non-SUSY SO(16)$\times$SO(16) string~\eqref{SO16_partition} by: 
\begin{center}
\includegraphics[scale=1]{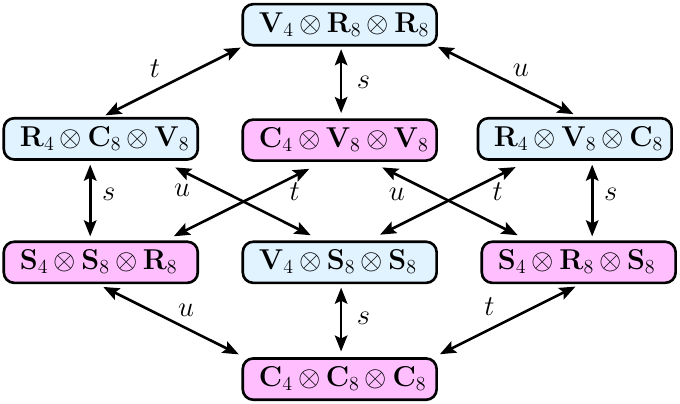}
\end{center}
Flipping the spin structure $s$ now not only exchanges the {\bf V}$_4$, {\bf R}$_4$  and the {\bf C}$_4$, {\bf S}$_4$ lattices, but at the same time the internal lattices get modified. 
This means that the bosons and fermions, that are related to each other by $s$--flips, live in different SO(16)$\times$SO(16) representations. Similarly, the spin structures $t$ and $u$ do not only change the gauge representations, but also the target-space statistics.

This diagram visualizes the symmetries between whole towers of string states; different mass states are mapped to each other. To see what remains of this structure at the massless level, we redraw the diagram, but this time only indicate the 10D charged massless spectrum: 
\begin{figure}[h!]
\includegraphics[scale=1]{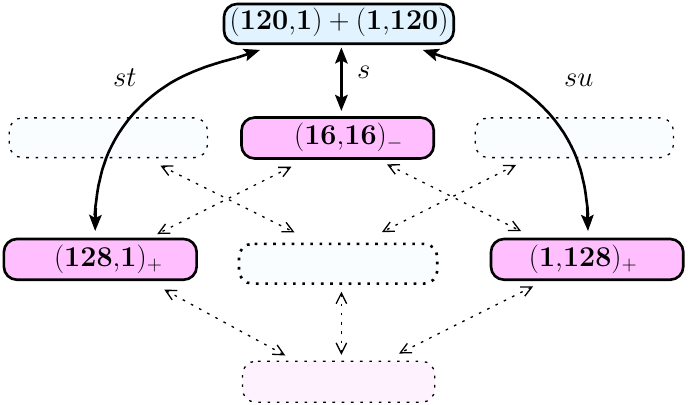}
\caption{\label{fg:MasslessSpinFlips} 
Spin-structure flips at the charged massless level}
\end{figure}

\noindent 
This diagram suggests that the SO(16)$\times$SO(16) gauge fields are related to the bi-fundamental $(\rep{16},\rep{16})$ fermions by the $s$-flip. Moreover, simultaneously flipping the spin structures, $s$ and $t$, or, $s$ and $u$, seems to relate the gauge fields to the fermions in the spinor representations of either SO(16) gauge group factor.

\subsection{B.\ Supersymmetry-breaking twists}

Next, we recall that the non-SUSY SO(16)$\times$SO(16) theory can be obtained from both  SO(32) and E$_8\times$E$_8$ theories by applying appropriate \cancel{SUSY} twists~\cite{Dixon:1986iz,Font:2002pq,Blaszczyk:2015zta}: 
\begin{center} 
\includegraphics[scale=1]{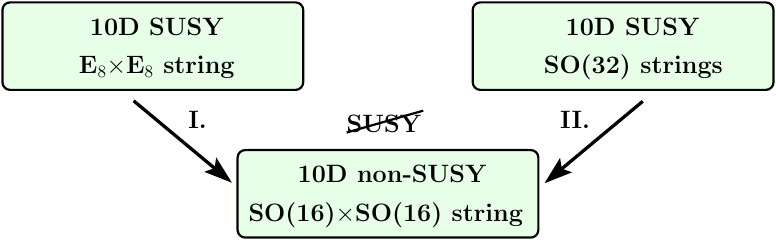}
\end{center} 
In fact, one can understand the full massless spectrum of the SO(16)$\times$SO(16) string as untwisted sectors of these \cancel{SUSY} twists only. 
In the redrawn diagram below we indicate which fields of the E$_8\times$E$_8$ and SO(32) super-Yang-Mills (SYM) theories survive (even) or disappear (odd) under the \cancel{SUSY} twists: 
\begin{center} 
\includegraphics[scale=1.75]{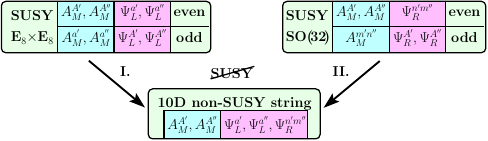}
\end{center} 
Via both routes the SO(16)$\times$SO(16) gauge fields survive as they are always even. In addition, the fermions in the spinor representations survive the \cancel{SUSY} projection of the E$_8\times$E$_8$ theory, while the bi-fundamental fermions survives the \cancel{SUSY} projection of the SO(32) theory.

\section{Twisting 4D N=1 superspace}

Next we want to develop a convenient formalism to describe such \cancel{SUSY} twists in a general context and shed light on the fermionic symmetries alluded above. To this end we turn to the N=1 superfield formalism in 4D (following the conventions of~\cite{wess1}). The dynamical components of the vector multiplets are $V = (A_\gm, \gl_\ga)$ and of chiral superfields $\gF = (z,\gps_\ga)$ using 4D Minkowski indices $\gm = 0,1,2,3$ and Weyl spinor indices $\ga, \dga = 1,2$. 
The chiral super gauge parameters $\gL = (\ga,\gr_\ga)$ of 
the 4D N=1 super gauge transformations, 
\equ{ \label{SuperGauge} 
e^{2V} \ra e^{2\bgL} \, e^{2V} \, e^{2\gL}~, 
\qquad 
\gF  \ra \varrho(e^{-2\gL})\,  \gF~, 
}
take values in the Lie algebra $\frak{g}$. The group homomorphism $\varrho$ defines the representation $R$ in which the chiral superfields $\gF$ live.

The SUSY breaking twist can be represented on N=1 4D superspace $(x^\gm,\gth^\ga,\bgth^\dga)$ as a reflection of the Grassmann coordinates~\cite{Blaszczyk:2015zta}: 
\equ{ \label{SUSYbreakingTwists}
\gth^\ga \ra  -\gth^\ga~, 
\qquad 
\bgth^\dga \ra -\bgth^\dga~, 
\qquad
x^\gm \ra x^\gm~. 
}
This corresponds to a particular $\Intr_2$ orbifold of superspace. 
The twist action can be extended to the superfields as: 
\equ{ \label{TwistBC}
\arry{lclcl}{
V(\gth) &\ra& V(-\gth) &=& Z\, V(\gth)\, Z~, 
\\[2ex]  
\gF(\gth) &\ra& \gF(-\gth) &=& \cZ\, \gF(\gth)~, 
}
}
where $Z$ is an unitary matrix that squares to the identity 
with a consistent choice for $\cZ = \varrho(Z)$.

The induced action on the super gauge parameters,  
\equ{ 
\arry{lclcl}{
\gL(\gth) &\ra& \gL(-\gth) &=& Z\, \gL(\gth)\, Z~, 
}
} 
splits the gauge algebra $\frak g$ in even $\frak g_+$ and odd $\frak g_-$ parts. Since 
\equ{
[ {\frak g}_+, {\frak g}_+ ] \subset {\frak g}_+~, 
\quad 
[ {\frak g}_+, {\frak g}_- ] \subset {\frak g}_-~, 
\quad 
[ {\frak g}_-, {\frak g}_- ] \subset {\frak g}_+~, 
}
the even part ${\frak g}_+ \subset {\frak g}$ is a subalgebra, which identifies the unbroken gauge interactions: The remaining gauge parameters $\ga$ and gauge fields $A_\gm$ reside in $\frak{g}_+$; the surviving gauginos live in $\frak g_-$.

Similarly, the representation $R$ is decomposed in even $R_+$ and odd $R_-$ parts: The surviving complex scalars $z$ reside in $R_+$, while the surviving fermions are in $R_-$. 
This can have serious consequences: Even if the original SUSY theory is anomaly-free, the fermionic spectrum after the \cancel{SUSY} projection will in general be anomalous. 
The non-SUSY heterotic string solves this problem in an elegant way: The untwisted fermionic sectors under the \cancel{SUSY} twists of the SUSY E$_8\times$E$_8$ or SO(32) theories are anomalous separately, but when combined an anomaly-free spectrum is obtained. 
Similar pairs of SUSY theories, that upon \cancel{SUSY} twists give the same bosonic spectrum and complementary fermionic spectra which are anomaly-free when combined, are expected to exist in general.

\subsection{Fermionic transformations} 

Also the fermionic parameters $\gr_\ga \in \frak{g}_-$ of the super gauge transformations~\eqref{SuperGauge} survive the \cancel{SUSY} twist. 
If we employ the Wess-Zumino (WZ) gauge: 
$
\smash{V| \stackrel{!}{=} 
D^2 V| \stackrel{!}{=} 0}$ 
and 
$\smash{\gch_\ga = \sfrac 1{\sqrt 2} 
D_\ga V| \stackrel{!}{=} 0}$,  they e.g.\ correspond to the following fermionic transformations: 
\equ{ \label{FermTrans} 
\gd \gch_\ga = \gr_\ga~, 
\qquad 
\gd \gl_\ga = - \sfrac1{\sqrt 2}\, \ge^{\dga\dgb} \gs^m_{\ga\dga}\, [A_m, \bgr_\dgb]~. 
}
For simplicity we have given only the non-trivial transformations. 
Note that the fermionic transformations~\eqref{FermTrans} do not preserve the WZ gauge.

This can be generalized to chiral superfields: To obtain non-trivial fermionic variations among their remaining components we need non-Abelian super gauge transformations 
before the \cancel{SUSY} twist with $\cZ\neq \pm \Id$.

 A related modification of SUSY was proposed in~\cite{Bars:1996xm}: There the SUSY generators were combined with some derivatives in additional internal directions.
 However, here we do not construct some intricate modification of SUSY, but rather we consider a substructure of the SUSY-gauge algebra defined by the fermionic parts of the super gauge transformations. For this reason our result is not in conflict with the Haag-Lopuszanski-Sohnius theorem~\cite{Haag:1974qh}. 
Fermionic symmetries evading this theorem were suggested for adjoint-QCD theories in the large $N$ limit where their S-matrices become trivial~\cite{Basar:2013sza,largeNQCD}.

\subsection{ Supergraphs on twisted superspace }

The methods of supergraphs (see e.g.~\cite{Buchbinder:1998qv}) can be applied immediately  to \cancel{SUSY}-twisted theories when one uses twist compatible superspace delta functions. (This is inspired by the construction of orbifold compatible delta functions in~\cite{GrootNibbelink:2005vi,Nibbelink:2006eq}.) 
They arise naturally by introducing chiral sources $J^a$ associated with $\gF^a$. Since they need to have the same twist properties~\eqref{SUSYbreakingTwists}, their functional derivatives, 
\equ{ 
 \frac{\gd J_{2}{}^a}{\gd J_{1}{}^b} = -\sfrac 14 \bD^2_2\, \big(\tgd_{21}\big)^a{}_b~, 
}
give rise to \cancel{SUSY}-twist compatible superspace delta functions:
\equ{ \label{TwistCompDelta} 
\big(\tgd_{21} \big)^a{}_b = \sfrac 12\, \gd^4_{21} \, 
\Big\{     
(\gth_2  - \gth_1)^4 \gd^a{}_b +  (\gth_2  +  \gth_1)^4\, \cZ^a{}_b
\Big\}~.  
}
Here, the subscripts $1, 2$ labels two different superspace coordinate systems, $\gd^4_{21} = \gd^4(x_2-x_1)$ is the 4D delta function and chirality of the functional derivatives is
 ensured by $\smash{-\sfrac 14 \bD^2}$.

\section{Examples of \cancel{SUSY}-twisted theories} 

\subsection{ 4D SU(2) super-Yang-Mills}

To illustrate this procedure we consider a 4D N=1 SU(2) SYM theory. The vector superfield $V$ is coupled to an adjoint chiral superfield $\gF  =(z, \gps_\ga, F)$ with the auxiliary field $F$. We define the \cancel{SUSY} twist to act as 
\equ{
\arry{lclcl}{ 
\gF(\gth) &\ra& \gF(-\gth) &=& \gs_3\, \gF(\gth)\, \gs_3~, 
}
}
and the same for $V$. To investigate which components survive, decompose e.g.\ the chiral superfield as 
\equ{ \label{SU2expansion} 
\gF = \gF_{+}^{\scriptscriptstyle (\boldsymbol 0)} \, \gs_3 + \gF_{-}^{\scriptscriptstyle (\boldsymbol \pm)}\, \gs_\pm~, 
}
using the basis $\gs_i = \gs_3, \gs_\pm$ and a sum over $\pm$ is implied. 
The unbroken U(1) charges, $\sfrac 12 \gs_3$,  
are indicated in brackets; when the U(1) charge is irrelevant we suppress the charges and interpret $\gF_-$ as a two-component vector:
\[ \renewcommand{\arraystretch}{1.2}
\arry{|l|c|c|c|}{
\hline 
\multicolumn{1}{|c|}{\textbf{Chiral}} &\multicolumn{2}{c|}{\textbf{Surviving}} & \textbf{U(1)} \\
\multicolumn{1}{|c|}{\textbf{superfield}} & \textbf{bosons} & \textbf{fermions} & \textbf{charge}  
\\ \hline\hline 
\rule{0pt}{12pt} \gF_+ = \big( \gF_{+}^{\scriptscriptstyle (\boldsymbol 0)} \big) & z, F & \text{--} & 0 
\\ 
\rule{0pt}{15pt} \gF_- = \big(\gF_{-}^{\scriptscriptstyle (\boldsymbol +)}, \gF_{-}^{\scriptscriptstyle (\boldsymbol -)}\big)
& \text{--} & 
 \big(\gps_\ga^{\scriptscriptstyle (\boldsymbol +)}, \gps_\ga^{\scriptscriptstyle (\boldsymbol -)} \big) & 
\big( \!+\!1,-1\! \big) 
\\[-3.3ex] &&& 
\\ \hline  
}
\]
The surviving bosonic and fermionic components are indicated in this table.
Contrary to the generic situation, in which the \cancel{SUSY} twist leads to an anomalous spectrum, in this case the resulting fermionic spectrum is anomaly free by itself.

Given the U(1) charge assignment the most general superpotential after the \cancel{SUSY} twist would be 
\equ{ \label{SuperP} 
W = t_+\, \gF_+ 
+ \sfrac 12\, (m_\pm + \gl_\pm\, \gF_+)\, \gF_\pm^2~, 
}
where $\gF_-^2 := \gF_{-}^{\scriptscriptstyle (\boldsymbol+)} \gF_{-}^{\scriptscriptstyle (\boldsymbol-)}$. Demanding invariance under the  fermionic symmetries induced by~\eqref{SuperGauge}:

\equ{
\gd \gps_\ga^{\scriptscriptstyle (\boldsymbol \pm)} = \pm 4\, \gr_\ga^{\scriptscriptstyle (\boldsymbol\pm)}\, z~, 
\quad 
\gd F = \pm 2 \, (\gr^{\scriptscriptstyle ( \boldsymbol \pm)})^{\ga} \, \ \gps_{\ga}^{\scriptscriptstyle (\boldsymbol \mp)}~, 
}
and $\gd z = 0$, one finds various coupling relations:
\equ{ \label{RestrictedCouplings} 
t_+ = 0~, 
\quad 
m_+ = m_-~, 
\quad 
3\, \gl_+ = 2\, \gl_-~. 
}
Hence, 
the construction of the non-SUSY theory with a fermionic symmetry is highly restrictive. 
In fact, if we start with an SU(2) invariant superpotential: 
\equ{ \label{SuperPSU2} 
W_\text{SU(2)} = \sfrac 12\, m\, \tr\, \gF^2~, 
}
we find that $\gl_\pm=0$ by inserting~\eqref{SU2expansion}, 
since there is no cubic superpotential for SU(2): 
$\gF^2 \sim \Id_2$ for SU(2) and the Pauli matrices $\gs_i$ are traceless.

\subsection{ 10D SUSY heterotic theories }

The \cancel{SUSY} twist of N=1 4D superspace can be extended to the 10D SYM theories~\cite{Marcus:1983wb,ArkaniHamed:2001tb} that arise from the heterotic strings: In these cases one has vector multiplets $V$ and three chiral superfields $\gF_i$, $i=1,2,3$, in the adjoint of either E$_8\times$E$_8$ or SO(32). Their components define gauge fields $A_M =(A_\gm, A_i = z_i)$ and gauginos $\gPs = (\gl, \gps_i)$ which are all functions of the full 10D spacetime coordinates, $x^M = (x^\mu, x^i, \bx^\ui)$. The 4D N=1 super gauge transformations are extended to 
\equ{ \label{SuperGauge10D} 
e^{2V} \ra e^{2\bgL} \, e^{2V} \, e^{2\gL}~, 
\quad 
\gF_i \ra e^{-2\gL}\, \big( \gF_i + \der_i \big)\, e^{2\gL}~. 
}

Hence, the fermionic components $\gr_\ga^{a'}$ of $\gL$ survive the \cancel{SUSY} twist of the E$\boldsymbol{_8\times}$E$\boldsymbol{_8}$ theory leading to fermionic transformations, like 
\equ{
\gd \gps_{i\,\ga}^{a'} = 2\, \der_i \gr_\ga^{a'} - (\gS_{A'}){}^{a'}{}_{b'}\, \gr^{b'}_\ga\, z_i^{A'}~, 
}
where $\gS_A$ are the SO(16) spin generators. 
These are precisely the variations suggested by the $st$-flip in FIG.~\ref{fg:MasslessSpinFlips}. Also to the other spin-structure flips we can associate fermionic transformations: The $su$-flip is obtained by replacing $' \ra ''$;  the $s$-flip corresponds to the fermionic transformations with $\gr^{m'n''}_\ga$ surviving the \cancel{SUSY} twist of the SO(32) theory.

\section{Quantum corrections}

In this final section we show that such \cancel{SUSY}-twisted theories still possess some remarkable quantum properties. To this end we consider a simple SUSY theory 
\equ{ \label{saction}
 S= \int \d^4 x 
 \Big\{ 
 \,\frac 12 \int \d^4 \gth\, \bgF_{\pm} \gF_{\pm} +  
\int \d^2\gth\, W + h.c.
\Big\}~, 
}  
with the superpotential~\eqref{SuperP} involving one even, $\gF_+$, and one odd, $\gF_-$, chiral superfield.

\subsection{One loop non-renormalization of the superpotential}

After the \cancel{SUSY} twist the resulting theory is non-SUSY (though for special spectrum and coupling choices accidental SUSY cancellations may occur),
nevertheless there are some remarkable non-renormalization results: As can be seen by a component analysis, the superpotential does not renormalize at one loop. 
For example, the $F$-term tadpole
\begin{center} 
\includegraphics[scale=1]{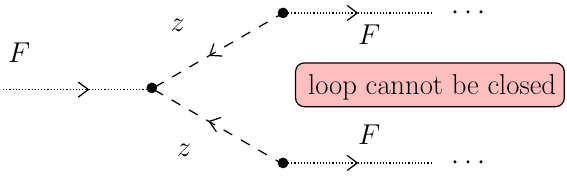}
\end{center}  
simply cannot be closed. Here the dotted lines refer to the auxiliary fields and the dashed lines to the scalar components of the chiral superfields $\gF_+$. 
Also in the SUSY context the one-loop non-renormalization is not a consequence of boson/fermion degeneracy but only of holomorphicity~\cite{Seiberg:1993vc}. Similar non-renormalization results were recently observed for higher dimensional operators in effective field theories~\cite{Elias-Miro:2014eia}.

However, beyond one-loop the non-SUSY superpotential interactions are not protected anymore. Two-loop diagrams, like  
\begin{center} 
\includegraphics[scale=0.95]{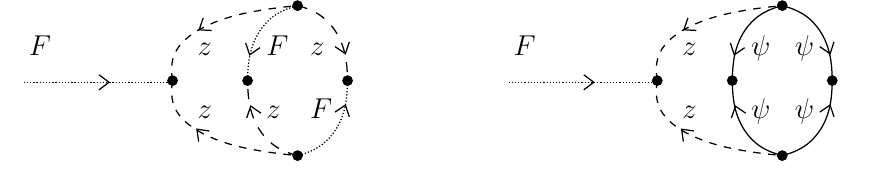}, 
\end{center}  
do not cancel, unless we invoke accidental SUSY cancellations referred to above.

\subsection{Chiral superfield tadpole}

At the one-loop explicit \cancel{SUSY} corrections are being generated. As an illustration we calculate the $z$-tadpole using \cancel{SUSY}-twist compatible supergraphs: 
\begin{equation} \label{QuadraticDivTadpole}
\raisebox{-0.51cm}{\includegraphics[scale=1.2]{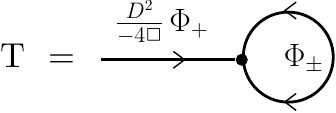}}
\end{equation}
(sum over $\pm$ implied).
The $ \sfrac {D^{2}} {-4 \Box}$ on the external line ensures that we compute the $z$-tadpole (without it we would reproduce the vanishing $F$-tadpole). Inserting the explicit expression for the \cancel{SUSY}-twist compatible superspace delta function~\eqref{TwistCompDelta} in the resulting expression, 
\equ{ \label{spurion}
T = 
\sfrac{1}{4}\, \gl_\pm\!
\int\!(\d^4 x\,\d^4 \gth)_{12}\,  
\Big[ \sfrac {D^2} {-4 \Box} \gF_{+}\Big]_1
\tgd_{12}\,
\Big[\sfrac { \overline{m}_{\pm}  } {\square - |m_{\pm}|^{2}}\, \sfrac{\bD^2}{-4} \Big]_2
\tgd_{12}~, 
\non   
} 
gives rise to two types of contributions: One involves the ordinary Grassmann delta function $(\gth_2-\gth_1)^4$ giving a vanishing contribution; while the other  $(\gth_2+\gth_1)^4$ leads to a spurion $\gth^{2}$ insertion after integrating out one set of Grassmann variables. Extracting the divergent part we obtain 
\equ{ \label{divquadratic}
T_\text{div} = 
\mp \sfrac12 \,  \gl_{\pm}  \overline{m}_{\pm}
 \int\! \frac{\d^4 q}{(2 \gp)^{4}}\, 
\frac { 1 } {q^{2} + |m_{\pm}|^{2}}
\int\!  \d^4 x \, z~. 
}
using that $\int\! \d^4 \gth~ \gth^2\, \sfrac {D^2} {-4 \Box} \gF_{+} = z$ and $\cZ_{\pm} = \pm 1$.  
For a partially accidental SUSY tuning of the couplings, $\gl_{+} \overline{m}_{+} = \gl_{-} \overline{m}_{-}$, the quadratic divergences cancel; when also $m_{-} = m_{+}$ the tadpole vanishes entirely.

\subsection{One-loop induced ``soft''-terms}

As demonstrated by the supergraph calculation above  \cancel{SUSY} effects can be induced. Such ``soft'' terms~\cite{Girardello:1981wz} can be parameterized using the spurion $\gth^2$ as 
\equ{
S_\text{``soft''} = \int\d^4x \d^2\gth~ \ \gth^2\, \tW~, 
}
where 
\(
\tW  = \widetilde{t}_+\, \gF_+ 
+ \sfrac 12\, (\widetilde{m}_\pm + \widetilde{\gl}_\pm\, \gF_+)\, \gF_\pm^2 
\)
has the same structure as the superpotential~\eqref{SuperP} but with different couplings. 
For the following reason we put the borrowed terminology of ``softness'' 
in quotation marks: Contrary to soft-terms in SUSY theories, tadpoles (like~\eqref{QuadraticDivTadpole}) may, in fact, be quadratically divergent 
for the class of non-SUSY theories under investigation.

\section{Acknowledgements}

We would like to thank Orestis Loukas and Patrick K.S.\ Vaudrevange for careful reading the manuscript.
Moreover, SGN would like to thank Xiao Liu of UESTC and Carlo Angelantonj of Turin University for their kind hospitality and useful discussions.

\bibliography{paper}

\end{document}